\documentclass[aps,showpacs]{revtex4}

\usepackage{graphicx}

\newcommand{\eqref}[1]{(\ref{#1})}
\renewcommand{\d}{\partial}
\newcommand{\nn}{\nonumber\\}

\newcommand{\ph}{\varphi}

\newcommand{\ep}{\varepsilon}

\newcommand{\MSbar}{{\ensuremath{\overline{\mathrm{MS}}}}}
\renewcommand{\c}[1]{{\cal{#1}}}

\begin{document}

\title{Renormalization and resummation in finite temperature field theories}

\author{A. Jakov\'ac}
\email{jakovac@planck.phy.bme.hu}
\affiliation{Hungarian Academy of Sciences and Budapest
  University of Technology and Economics Research Group `Theory
  of Condensed Matter'', H-1521 Budapest, Hungary}
\author{Zs. Sz{\'e}p}
\email{szepzs@achilles.elte.hu}
\affiliation{Research Group for Statistical Physics of the
  Hungarian Academy of Sciences, H-1117 Budapest, Hungary}

\begin{abstract}
  Resummation, ie. reorganization of perturbative series, can result
  in an inconsistent perturbation theory, unless the counterterms are
  reorganized in an appropriate way. In this paper two methods are
  presented for resummation of counterterms: one is a direct method
  where the necessary counterterms are constructed order by order; the
  other is a general one, based on renormalization group arguments. We
  demonstrate at one hand that, in mass independent schemes, mass
  resummation can be performed by gap equations renormalized
  \emph{prior to} the substitution of the resummed mass for its
  argument.  On the other hand it is shown that any
  (momentum-independent) form of mass and coupling constant
  resummation is compatible with renormalization, and one can
  explicitly construct the corresponding counterterms.
\end{abstract}
\pacs{11.10.Wx, 11.10.Gh}
\maketitle

\section{Introduction}

Resummation is often required in a perturbation theory to avoid
infrared (IR) divergences to appear at finite energy scales. Examples
are 1PI diagram resummation (Schwinger-Dyson equation) to cure the
on-shell singularity of the propagator, or daisy resummation
\cite{DolanJackiw} to resum IR divergences of massless theories at
finite temperature. We need leading (or higher order) logarithmic
resummation near a second order phase transition point
\cite{secondorder}, and the HTL resummation \cite{BrPis} to be able to
consistently define finite temperature observables in gauge theories.

Resummation means reorganization of the perturbative series in a way
that the diagrams that are most important from the point of view of IR
behavior, are taken into account first. However, IR importance does
not necessarily means UV importance as well, and so it can happen that
those (counterterm) diagrams that are needed to make the theory
finite, are shifted to a later stage in the resummation process. Then
we observe that lower order results are divergent meaning that the
resummed perturbation theory is \emph{inconsistent}.  Exceptions are
the 1PI and leading log resummations, where, by chance, diagrams of IR
and UV importance are the same, but other cases (daisy, super-daisy,
HTL etc.  resummations) suffer from this difficulty.  Therefore, in
these cases, the physically well-motivated resummation equations (gap
equations) are meaningless in principle, and one usually makes some ad
hoc assumption to obtain finite results.

Recently several papers were published that investigate this problem
in various physical situations \cite{recent}. In the present paper we
try to go around the question of how to construct countertem diagrams
for a given \emph{momentum independent} resummation. We try to give
two points of view on the problem: one is a direct method where we
make explicit resummation for the counterterm diagrams, induced by the
resummation of the ``normal'' diagrams (Section~\ref{sec:massresum}).
The relevant counterterm diagrams can be constructed in each
perturbative order, and finally we give an explicit form for the
resummed Lagrangian. To be more specific, we suggest, in $\Phi^4$
theory, the following reorganization of the mass terms in the
Lagrangian in \MSbar\ scheme to regularize the mass resummation:
\begin{equation}
  \label{reorganization}
  -\c L_\mathrm{mass} = \frac{M_T^2}2 \ph^2 +
  \underbrace{\left(-\frac{\Delta M_T^2}2 \ph^2 + \frac{z_\MSbar
        M_T^2}2 \ph^2\right)}_{\mbox{\small one loop}} +
  \underbrace{\left(-\frac{z_\MSbar\Delta M_T^2}2\ph^2
    \right)}_{\mbox{\small two loops}},
\end{equation}
where $z_\MSbar$ is the multiplicative mass renormalization factor in
\MSbar\ scheme, and $M_T^2 = m^2 + \Delta M_T^2$. This proposal is a
generalization of the method of Banerjee and Mallik \cite{Banerjee}.
At one hand this expression is equivalent with $(m^2+z_\MSbar
m^2)\ph^2/2$, so non-perturbatively we have the same physics. On the
other hand we will prove that, if we take into account the
under-braced terms first in one, and two loop levels, respectively, we
obtain a mass resummed perturbation theory that consistently removes
all the divergencies at any order.

One of the most important corollary of these investigations is the
following rule of thumb: we can renormalize the gap equations for mass
resummation by \emph{first renormalizing the diagrams} in an
appropriate mass independent physical scheme (eg. \MSbar), and
\emph{afterwards substituting the resummed parameters} in the finite
expressions. As it will be seen later on, although this is not the
most generic renormalization method, still it is a possible and
consistent way to deal with UV divergences. This strategy was already
used in Ref.~\cite{JPSS}, in the study of phase diagram of the
quark-meson model.

The other point of view treats resummed perturbation theory as a
different renormalization scheme (Section~\ref{sec:RG}). We call this
scheme, in which we impose environment ($T,\,\mu$ or even time)
dependent renormalization conditions as \emph{resummation (RS) scheme}.
This is not a physical choice, since changing the environment results
in changing of the renormalization scheme. In order to have a
meaningful result we have to relate the RS scheme and a bona
fide renormalization scheme, eg. \MSbar. This means constructing
relations between the parameters of the renormalized Lagrangians of
the two schemes which formally appear as \emph{gap equations}.

In several applications we need the inverse procedure: one has a
physically motivated gap equation, and one would like to know the
corresponding structure of counterterms. In this approach we consider
the gap equations as relations between the \MSbar\ and some
RS scheme. Expanding the gap equation in terms of the
coupling constant we can read off the finite difference that has to be
added to the \MSbar\ counterterm to obtain the RS scheme
counterterm. Thus the RS scheme is defined, and so we can use
perturbation theory to calculate any observables we need. If the
choice of the gap equations was good enough, we will observe
cancellations between the ``normal'' and counterterm diagrams. The
diagrams that survive cancellation are not part of the resummation,
and they can be used eventually to improve the resummation process.

The paper ends with conclusions and an outlook
(Section~\ref{sec:conclusion}).

\section{Mass resummation in \MSbar\ scheme}
\label{sec:massresum}

Our basic model is the $\Phi^4$ model where the Lagrangian reads in
terms of renormalized couplings and fields as
\begin{equation}
  \c L = \frac12 (\d_\mu\ph)^2 - \frac{m^2}2\ph^2
  -\frac\lambda{24}\ph^4 +\frac{Z^2}2 (\d_\mu\ph)^2 - \frac{\delta
    m^2}2\ph^2 -\frac{\delta\lambda}{24}\ph^4.
\end{equation}
As it is well known \cite{DolanJackiw}, in this model perturbation
theory becomes unreliable at high temperatures, because of the
presence of diagrams giving $(T/m)^N$ contribution. This (IR) problem
can be cured by summing all the dominant tadpole contribution of the
theory (``daisy diagrams''). By this procedure one effectively
replaces the mass $m^2$ by the resummed mass $M_T^2 = m^2+\Delta
M_T^2$ (where, in this theory, $\Delta M_T^2 =\frac\lambda{24} T^2$ to
leading order), making the IR problem disappear (apart from the
phase transition point \cite{secondorder}).

Since the daisy resummation does nothing else than replacing the mass
with a different mass, it is equivalent with the ``thermal
counterterm'' procedure \cite{thermalmass}. We add to and subtract
from the Lagrangian the same, temperature dependent term; the relevant
part of the Lagrangian therefore reads
\begin{equation}
  \label{eq:thermmass}
  \frac{m^2}2\ph^2 +\frac{\delta m^2}2\ph^2 = \frac{m^2+\Delta
    M_T^2}2\ph^2 - \frac{\Delta M_T^2}2\ph^2 +\frac{\delta m^2}2\ph^2,
\end{equation}
As usual, the added term is treated as a part of the free Lagrangian
while the subtracted term, on the other hand, is taken into account
one loop later as a (thermal) counterterm. By doing this we
reinterpreted the perturbation theory, ie. we defined a resummation.
The value of $\Delta M_T^2$ can be chosen to be the tadpole value, but
its ``best value'' can be found out from the condition that at one
loop order the self-energy is zero
\begin{equation}
  \label{gapequation}
  \Pi(p=0,M_T^2) - \Delta M_T^2 =0.
\end{equation}
This is a gap equation \cite{Andersen, Buchmuller} which is to be
solved to find $\Delta M_T^2$.

It turns out, however, that this naive definition is not consistent
\cite{recent,Andersen,PatkosPetreczky} because of the mismatch in the
counterterm diagrams. To see this we calculate the one loop self
energy in the $\Phi^4$ theory
\begin{equation}
  \Pi(p=0,M_T^2) = m^2+\Delta M_T^2 + \frac\lambda2 T_B(m^2+\Delta
  M_T^2) - \Delta M_T^2 + \delta m^2,
\end{equation}
where $T_B$ is the bosonic tadpole function; in dimensional
regularization it reads:
\begin{equation}
  \label{TB}
  T_B(M) = \frac{M^2}{16\pi^2} \left[-\frac1\ep +\gamma_E-1
    +\ln\frac{M^2}{4\pi\mu^2} \right] +
  \frac1{2\pi^2}\int\limits_M^\infty\!
  d\omega\,\sqrt{\omega^2-M^2}\,n(\omega),
\end{equation}
where $n(\omega)$ is the Bose-Einstein distribution. 

We should write $T_B$ with argument $M=M_T$, and that is what creates
problem here. We, namely, already renormalized the theory at zero
temperature, say, in \MSbar\ scheme. That fixes the one-loop mass
counterterm as
\begin{equation}
  \label{MSbardeltam2}
  \delta m_1^2 = -m^2 \frac\lambda{32\pi^2} \left[-\frac1\ep
    +\gamma_E-1 - \ln 4\pi \right].
\end{equation}
So we have a divergency $\sim -M_T^2/\ep$ and a counterterm $m^2/\ep$,
that do not cancel each other. There is an unbalanced divergency, and
so the RS scheme is inconsistent.

Since at the non-perturbative level (infinite order) the resummed
theory is equivalent with the original one (they have the same
Lagrangian), we expect that this divergence vanishes by higher loop
effects. It has been shown \cite{PatkosPetreczky} that in the next
order (two loop in case of mass) we indeed obtain contributions that
cancel this divergence -- however, other new unbalanced divergences
appear in this order. So finally the perturbation theory will not be
consistent at any finite loop level.

\subsection{Resummation of counterterm diagrams}

In order to make the theory consistent, we should find some method to
shift up to one loop level those two-loop diagrams that are necessary
to cancel the unbalanced divergences at one loop order: ie. we have to
reorganize (resum) the \textrm{counterterm diagrams}.  The idea is the
same as in the case of the thermal mass. We find the correction to the
counterterm, add it to and subtract it from the Lagrangian, but
classify them to belong to different loop orders. Let us denote the
necessary mass term by $\delta m_T^2$ and call it \emph{compensating
  counterterm}. So, instead of \eqref{eq:thermmass}, we use the
following mass terms in the Lagrangian:
\begin{equation}
  \label{eq:corrthermmass}
  \underbrace{\frac{m^2+\Delta M_T^2}2\ph^2}_{\mbox{tree\ level}}+
  \underbrace{\left(- \frac{\Delta M_T^2}2\ph^2 + \frac{ \delta m^2
        + \delta m_T^2}2\ph^2\right)}_{\mbox{one-loop\ level}} +
  \underbrace{\left(-\frac{\delta m_T^2}2
      \ph^2\right)}_{\mbox{two-loop\ level}}. 
\end{equation}
In perturbation theory the first term is taken into account in the
propagator, the second and third are taken into account first at one
loop level, while the last term first contributes at two-loop level.
$\delta m_T^2$ has to be determined order by order in a way that it
cancels the remaining divergence of the physical observable at a given
order. Although it is temperature dependent and also divergent in the
present example, its value is irrelevant from the point of view of the
consistency of the complete theory, since the Lagrangian is the same
as the original one, it does not depend on the value of $\delta
m_T^2$.

Using the proposed decomposition of the Lagrangian corresponding to
\eqref{eq:corrthermmass} let us compute the resummed and renormalized
self-energy of the $\Phi^4$ model.

\subsubsection{One loop self-energy}

At one loop level the self-energy reads with this new counterterm
structure as
\begin{equation}
  \label{1loopSE}
  \Pi_\mathrm{one-loop} = M_T^2 + \frac\lambda2 T_B(M_T^2) - \Delta
  M_T^2 + \delta m_1^2 + \delta m_{T1}^2.
\end{equation}
$\delta m_1^2$ is defined in \eqref{MSbardeltam2}; it cancels a part
of the divergence hidden in $T_B$, the rest has to be canceled by
$\delta m_{T1}^2$. We can choose it in analogy with the \MSbar\ form
\eqref{MSbardeltam2}
\begin{equation}
  \label{eq:deltamT2}
  \delta m_{T1}^2 = -\Delta M_T^2 \frac\lambda{32\pi^2}
  \left[-\frac1\ep +\gamma_E-1  - \ln 4\pi \right].
\end{equation}
In principle we could add an arbitrary finite term to this form. This
would modify the value of $\Delta M_T^2$, and finally would lead to a
different resummation process. In the language of Section~\ref{sec:RG}
it would correspond to a different renormalization scheme in the zero
and finite temperature parts. For further details
cf. Section~\ref{sec:RG}.

With this choice, using the form of $T_B$ from \eqref{TB} we find
\begin{equation}
  \label{1loopSE_result}
  \Pi_\mathrm{one-loop} = m^2 +\frac\lambda2 T_B^{\MSbar}(M_T^2),
\end{equation}
where 
\begin{equation}
  T_B^\MSbar(M^2) = \frac{M^2}{16\pi^2} \ln\frac{M^2}{\mu^2} +
  \frac1{2\pi^2} \int \limits_{M}^\infty\!d\omega\,
  \sqrt{\omega^2-M^2}\, n(\omega)
\end{equation}
is the \MSbar\ renormalized tadpole diagram. 

We could obtain this result by first renormalizing the divergent
diagrams in \MSbar\ scheme, and afterwards by substituting the
resummed mass into the finite expression.

\subsubsection{Two loop order}

Now let us turn to the determination of the self-energy at two loop
level. The contribution from two-loop level diagrams is represented
symbolically in the following way:
\begin{equation}
  \Delta \Pi_\mathrm{two-loop} =\frac{\lambda^2}6\;
  \raisebox{-0.3cm}{\includegraphics[height=0.8cm]{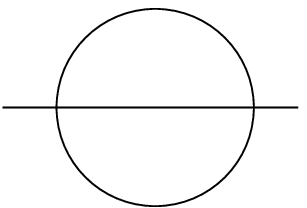}} +
  \frac{\lambda^2}4\;
  \raisebox{-0.25cm}{\includegraphics[height=1.6cm]{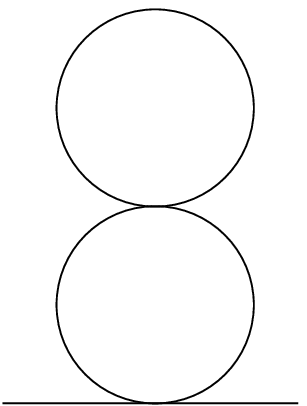}}
  + \frac{\delta\lambda_1}2
  \;\raisebox{-0.25cm}{\includegraphics[height=0.8cm]{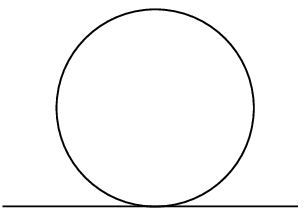}}
  + \frac\lambda2
  \;\raisebox{-0.25cm}{\includegraphics[height=0.85cm]{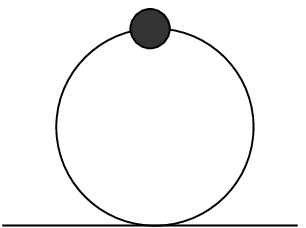}}
  -\delta m_{T1}^2 + \delta m_2^2 + \delta m_{T2}^2 - \delta
  Z_2 p^2.
\end{equation}
The propagators have the squared mass $M_T^2$, in this resummed
theory. The value of the diagrams in dimensional regularization,
writing out explicitly only their divergent part, are:
\begin{eqnarray}
  && \raisebox{-0.25cm}{\includegraphics[height=0.8cm]{tadpole.eps}} =
  \frac{M_T^2}{16\pi^2}\left(-\frac1\ep+\gamma_E-1+\ln\frac{M_T^2}
    {4\pi\mu^2} \right) + T_B^{T\neq0},\nn
  && \raisebox{-0.25cm}{\includegraphics[height=0.85cm]{fish.eps}} =
  (-\Delta M_T^2 + \delta m_1^2 + \delta m_{T1}^2) \left[
    \frac1{16\pi^2}\left(-\frac1\ep +\gamma_E +\ln\frac{M_T^2}
      {4\pi\mu^2} \right) + I_\mathrm{fish}^{T\neq0}\right],\nn
  &&\raisebox{-0.3cm}{\includegraphics[height=0.8cm]{setsun.eps}} =
  \frac{3M_T^2}{(4\pi)^4}\left[\frac1{2\ep^2}-\frac1\ep\left(\gamma_E
      -\frac32 +\ln\frac{M_T^2}{4\pi\mu^2}\right)\right] -
  \frac{p^2}{4(4\pi)^4} \frac1\ep - \frac3{16\pi^2}\,T_B^{T\neq0}\,
  \frac1\ep \;+\;I_\mathrm{setsun}^{finite}\nn 
  && \raisebox{-0.4cm}{
    \includegraphics[height=1.6cm]{doublescoop.eps}} =
  \frac{M_T^2}{(4\pi)^4} \left[\frac1{\ep^2}
    -\frac1\ep\left(2\gamma_E-1+2\ln\frac{M_T^2}{4\pi\mu^2}\right)
  \right] - \frac1{16\pi^2} \frac1\ep\left(T_B^{T\neq0} + M_T^2
    I_\mathrm{fish}^{T\neq0}\right) \;+\;I_\mathrm{dsc}^{finite}.
\end{eqnarray}
Since we do resummation in the \MSbar\ scheme, the value of $
\delta\lambda_1,\,\delta m^2_2$ and $\delta Z_2$ counterterms are
fixed, we cannot modify them. $\delta m^2_2 = z_{m2} m^2$, where
\begin{equation}
  \label{eq:zm2}
  z_{m2} = \frac{\lambda^2}{2(4\pi)^4}\left(\frac1{\ep^2} -
    (1-\ln4\pi)\frac1\ep\right)+\mathrm{finite}.
\end{equation}
The other counterterms read as
\begin{equation}
  \delta\lambda_1 = -\frac{3\lambda^2}{32\pi^2}\left(-\frac1\ep
    +\gamma_E -\ln4\pi\right),\qquad \delta Z_2= \frac1{4(4\pi)^4} \,
    \frac1\ep.
\end{equation}
Using these values the divergent part is still not canceled, what
remains is:
\begin{equation}
  \label{deltapi2div}
  \Delta \Pi_\mathrm{two-loop}^{div} = -
  \frac{\lambda^2\Delta M_T^2}{2(4\pi)^4}\left(\frac1{\ep^2} -
    (1-\ln4\pi)\frac1\ep\right) +\delta m_{T2}^2.
\end{equation}
To cancel this divergency we can choose $\delta m_{T2}^2 = z_{m2}
\Delta M_T^2$, although, of course, an arbitrary finite part could be
added to this expression.

With this choice, however, we get further support to the practical
observation of the one loop calculation: we first do renormalization
in \MSbar\ scheme, and substitute the resummation mass in the finite
expressions. It should be clear, too, that we could add finite
terms for $\delta m_{T1}^2$ and $\delta m_{T2}^2$, and then this
statement would not be true any more. Thus this is just a --
comfortable -- possibility.

\subsection{Higher loop orders}
\label{sec:higherloop}

One can conjecture that this feature persists at higher loop orders,
since in \MSbar\ scheme the overall divergency of a diagram always can
be written as $-z_mM^2$ with some mass-independent $z_m$ factor. Then,
by choosing $\delta m^2= z_m m^2$ and $\delta m_T^2= z_m \Delta
M_T^2$, the overall divergency disappears. This should also be true
for any other mass-independent schemes. We can, therefore, put forward
a proposition:

\begin{list}{\textbf{Proposition:}}
\item If in a renormalization scheme the zero temperature counterterm
  is $\delta m^2=z_mm^2$ and $z_m$ is mass-independent, then at finite
  temperature the theory with thermal counterterm $\Delta M_T^2$ and
  compensating counterterm $\delta m_T^2 = z_m \Delta M_T^2$ yields
  finite result at each order of the perturbation theory.
\end{list}

\begin{list}{\emph{Proof}:}{}
\item Let us consider a 2-point diagram $\c D(m^2)$ in the original
  (not resummed) theory that has an overall divergence, and all the
  subdivergences were consistently subtracted. Then, if the theory is
  renormalizable, the divergence of the diagram should be proportional
  to $m^2$; the proportionality constant can be obtained as $\d_{m^2}
  \c D(m^2)|_\mathrm{div}$. Since the divergence is canceled by the
  counterterm $\delta m^2=z_mm^2$, the divergent part should be equal
  to an appropriate part from the mass renormalization factor $z_m$,
  which is mass independent
  \begin{equation}
    \label{Dz}
    \d_{m^2} \c D|_\mathrm{div} = -z_{\c D}.
  \end{equation}
  
  In the resummed theory we have two effects. One is the substitution
  of the mass term by the resummed mass: $m^2\to m^2+\Delta M_T^2$.
  The divergent parts of the self energy diagrams, in our
  mass-independent scheme, are proportional to this new mass term. But
  we have also prepared the corresponding counterterm, since the third
  term of \eqref{eq:corrthermmass} reads $\delta m^2+\delta m_T^2 =
  z_m (m^2+\Delta M_T^2)$. So this part of the divergences disappear
  in the same way as in the unresummed case.
  
  The other effect is the $-\Delta M_T^2$ insertion coming from the
  second (thermal) counterterm of \eqref{eq:corrthermmass}. The sum of
  the single mass insertion diagrams is just $-\Delta M_T^2\d_{m^2} \c
  D(m^2)$, because at each propagator $G^2(p,m^2) = \d_{m^2}G(p,m^2)$
  (it is evident for Feynman propagators, at finite temperature it can
  be seen using imaginary time propagators). Its divergent part can be
  written, using \eqref{Dz}, as $\Delta M_T^2 z_{\c D}$. This is
  exactly canceled by the corresponding part of the compensating
  counterterm at this order: $-\delta m_T^2 = -\Delta M_T^2 z_m$.

  So at each diagram the overall divergences are canceled by the
  proposed counterterms. Using BPHZ argumentation, this is enough to
  make the complete theory finite.
\end{list}

\section{Renormalization group analysis}
\label{sec:RG}

Now we have a consistent method for how to reorganize the perturbation
theory according to a given (mass) resummation in such a way as to
have finite results in all orders in \MSbar\ scheme. In order to
generalize the result so as to be able to treat other schemes (not
just \MSbar), or environment dependence of other couplings (cf. vertex
resummation), we change our viewpoint, and we rephrase the complete
resummation procedure using the renormalization group (RG)
argumentation: this will be the topic of the present section.

First we demonstrate at one loop level that we can obtain the results
of the previous section using a special renormalization scheme. We
define the mass counterterm of the new scheme as
\begin{equation}
  \label{newscheme}
  \delta m^2 = \delta m_\MSbar^2 - \Delta M^2,
\end{equation}
where $\delta m_\MSbar^2=z_\MSbar m^2$ denotes the \MSbar\ counterterm
and $\Delta M^2$ is an arbitrary, finite expression. The other
counterterms are the same as in \MSbar\ scheme. Then the one loop self
energy reads
\begin{equation}
  \label{newschemeselfen}
  \Pi = m^2 +\frac\lambda2 T_B(m^2) + \delta m_\MSbar^2 - \Delta M^2.
\end{equation}
This expression differs in a finite term from the corresponding
\MSbar\ expression. At one hand it assures finiteness, on the other
hand this means that the new mass parameter must be different from the
\MSbar\ mass parameter in order to describe the same physics.  In
fact, applying the ideas of the renormalization theory \cite{Collins}
to the present case, there must be a choice $m^2(m_\MSbar^2)$ for
which we have the same value \emph{for all observables} in the two
schemes at a given order. This function can be read off from the
requirement that the bare parameters are the same in the two schemes
\cite{Collins}:
\begin{equation}
  \label{matchnewscheme}
  m^2 +\delta m^2 = m_\MSbar^2 + \delta
  m_\MSbar^2,\qquad\Rightarrow\qquad m^2 = m_\MSbar^2 + \Delta M^2.
\end{equation}
Then the counterterm of the new scheme \eqref{newscheme} can be
written as
\begin{equation}
  \delta m^2 =  - \Delta M^2 + z_\MSbar m_\MSbar^2 + z_\MSbar \Delta
  M^2.
\end{equation}
This is the result of the Proposition of subsection
\ref{sec:higherloop} applied to the one loop case.

For the definition of the new scheme we need to specify a value for
$\Delta M^2$.  We may, for example, choose the mass shell scheme,
where the full self-energy reads $\Pi(p=0) = m^2$. From
\eqref{newschemeselfen} and \eqref{matchnewscheme} we find for the
value of $\Delta M^2$
\begin{equation}
  \frac\lambda2 T_B^\MSbar(m_\MSbar^2 + \Delta M^2) - \Delta M^2=0,
\end{equation}
where $T_B^\MSbar$ now means the tadpole diagram, renormalized in the
$\MSbar$ scheme. This equation is the renormalized gap equation for
the mass.

This analysis suggests that the mass resummation is equivalent with
the choice of a new renormalization scheme. Using the relation of the
masses of the two schemes, we can compute finite observables that can
be interpreted as the result of the resummed \MSbar\ scheme. The mass
relation, together with the definition of the new scheme yields finite
gap equation for the new (resummed) mass, which is the usual gap
equation in case of the mass shell scheme.

\subsection{Generalization}

Motivated by the previous subsection, we can start to build up the
resummation strategy from the point of view of renormalization
schemes. Resummation itself means that some higher order diagrams are
taken into account at lower levels with the consequence that these
diagrams are missing from the higher order calculation. So, at least
for the mass and coupling constant resummation, resummation is
equivalent with a scheme that cancels these diagrams with a proper
choice of the finite part of the counterterms. Since we play only with
the finite part of the counterterms, we will still have a consistent
scheme. So this new scheme -- the \emph{resummation (RS) scheme} --
differs from a general scheme (eg. \MSbar) in that a certain set of
diagrams is missing from the calculation.

The RS scheme is a bona fide scheme while we keep the
counterterms fixed: for example we can equally well use the MS and the
\MSbar\ schemes. However, at finite temperature, the finite parts of
the counterterms will depend on temperature, and so the resummation
scheme itself varies with temperature. In a more general case
RS scheme can depend on the complete environment
\cite{envfriend}: beyond the temperature on chemical potential,
background condensates, time, etc.  Since results in different schemes
have different physical interpretation, RS scheme cannot directly
describe the effects of the environment variation. Example is the mass
shell scheme where $\Pi(p=0)=m^2$, and we have no explicit temperature
dependence. To be able to draw physical consequences, we have to
project the results to a common reference scheme.

Technically what we should do is to relate the RS scheme to a
physical renormalization scheme -- for concreteness we will use the
\MSbar\ scheme. Since both of them are mathematically correct schemes,
in case of renormalizable theories, this can be accomplished by
changing the values of renormalized parameters of the Lagrangian
\cite{Collins} (including wave function renormalization).  In $\Phi^4$
theory there must exist environment-dependent parameters
\begin{equation}
  \label{eq:relation}
  m=m(m_\MSbar,\lambda_\MSbar,T),\quad \lambda = \lambda(m_\MSbar,
  \lambda_\MSbar, T),\quad \zeta=\zeta(m_\MSbar,\lambda_\MSbar,T)  
\end{equation}
such as for any $n$-point function with momenta $\{p_i\}$ we have
\begin{equation}
  G^{(n)}_\MSbar(p_i;m_\MSbar,\lambda_\MSbar) = \zeta^n
  G^{(n)}_\mathrm{resum}(p_i;m,\lambda).
\end{equation}
In perturbation theory the equality holds only up to the computed
order; in fact the difference is the effect of resummation. The value
of the ``running'' parameters can be computed using the fact that the
bare quantities are independent of the scheme \cite{Collins}. In this
way we can use the RS scheme for perturbation theory,
enjoying its good IR convergence, and still have results in a physical
renormalization scheme (eg. \MSbar), where also the parameters of the
theory can be obtained from the usual renormalization conditions.

Sometimes we can explicitly define the new scheme (eg. the mass shell
scheme for mass resummation), but more often we just have a guess for
the resummed parameters, and we hope that it performs some kind of
resummation. That is we start with explicit functions of the form of
\eqref{eq:relation} for the resummed mass and coupling constant (and
eventually, wave function renormalization). According to the line of
thought above these relations can be interpreted as generators of a
RS scheme: we expand them in power series in $\lambda_\MSbar$
\begin{equation}
  m^2 = m_\MSbar^2 + \sum\limits_{n=1}^\infty \lambda_\MSbar^n \Delta
  M_n^2,\quad \lambda = \lambda_\MSbar + \sum\limits_{n=1}^\infty
  \lambda_\MSbar^n \Delta \lambda_n,\quad \zeta = 1 +
  \sum\limits_{n=1}^\infty \lambda_\MSbar^n \Delta Z_n, 
\end{equation}
and we interpret $\Delta M_n^2$ as the $n$th order finite mass
counterterm that has to be added to the \MSbar\ counterterm;
similarly, $\Delta \lambda_n$ is the finite coupling constant
counterterm, $\Delta Z_n$ is the finite wave function renormalization
counterterm. Having defined the counterterms, the RS scheme
is appropriate to do perturbation theory, and the results can be
related to the \MSbar\ scheme by the same relations
\eqref{eq:relation}. If our guess was good enough we will observe
cancellation between ``normal'' and counterterm diagrams at higher
orders. The main message, however, is that, independently of whether a
choice represents a physically meaningful resummation or not,
\emph{any choice} leads to a finite, consistent perturbation theory.

So as a possible strategy for resummation, we can do the following: we
perform regularized perturbation theory in a generic scheme, keeping
counterterms free, at a certain loop order. The divergent part of the
counterterms are then fixed, as usual, to cancel the UV divergences of
the diagrams, while the finite parts are fixed to cancel the IR
dangerous parts. So we defined a scheme, which is, however, only a RS
scheme, since the counterterms can depend on the environment. We
determine its relation to the \MSbar\ scheme of the form
\eqref{eq:relation} by requiring that the bare parameters should be
the same in the two schemes. This relation, by construction,
corresponds to a resummation up to this perturbative order, and we
should have some plausible guess to choose the higher order terms,
hoping that it still performs some kind of resummation. Different
choices, of course, correspond to different resummations. This
procedure can be repeated at each perturbative order, and so the
resummation process can be improved.

\section{Coupling constant and mass resummation}

Using the ideas of the previous subsection, in this section we perform
both mass and coupling constant resummation. This latter is important
to describe the ``softening'' of the theory close to a phase
transition point \cite{secondorder}, and also to consistently describe
vertex resummation \cite{Buchmullervertex}.

At one loop level we can start from the effective potential
\begin{equation}
  V = \frac12 m^2\Phi^2 +\frac\lambda{24}\Phi^4 +
  \Delta V(M^2) + \frac12 \delta m^2\Phi^2 +\frac{\delta
    \lambda}{24}\Phi^4,
\end{equation}
where $M^2=m^2+\frac\lambda2\Phi^2$. The effective, background
dependent mass and coupling constant then reads
\begin{eqnarray}
  \label{eq:V''}
  && \frac{d^2V}{d\Phi^2} = M^2 + \delta m^2 +
  \frac{\delta\lambda}2\Phi^2 + \lambda \Delta V'(M^2) +
  \lambda^2\Phi^2 \Delta V''(M^2), \\
  && \frac{d^4V}{d\Phi^4} = \lambda +\delta \lambda + 3\lambda^2
  \Delta V''(M^2) + \c O(\lambda^3 \Phi^2).
\end{eqnarray}
A possible way of doing coupling constant resummation that we make
vanish all of the terms below $\c O(\lambda^3 \Phi^2)$. That means:
\begin{equation}
  \delta \lambda = - 3\lambda^2 \Delta V''(M^2).
\end{equation}
For the mass resummation we demand that $V''=M^2$; with the $\delta
\lambda$ value defined above we obtain
\begin{equation}
  \delta m^2 = -\lambda \Delta V'(M^2) + \frac12\lambda^2\Phi^2 \Delta
  V''(M^2).
\end{equation}
This counterterm depends on the background only at $\c O(\lambda^3)$
order. If $\Delta V$ depends on the environment, these two
counterterms define an environment dependent RS scheme.

If we denote by $\Delta V'_\mathrm{div}$ and $\Delta V''_\mathrm{div}$
the divergent parts according to the \MSbar\ scheme, and by $\Delta
V'_\MSbar=\Delta V'-\Delta V'_\mathrm{div}$ and $\Delta
V''_\MSbar=\Delta V''-\Delta V''_\mathrm{div}$ the finite
parts\footnote{In \MSbar\ scheme $(\Delta V'_\mathrm{div})'\neq \Delta
  V''_\mathrm{div}$ and so $(\Delta V'_\MSbar)'\neq \Delta
  V''_\MSbar$}, then we can write
\begin{eqnarray}
  &&\delta \lambda_\MSbar = - 3\lambda^2 \Delta
  V''_\mathrm{div}(M^2)\nn 
  &&\delta m^2_\MSbar = -\lambda \Delta V'_\mathrm{div}(M^2) +
  \frac12\lambda^2\Phi^2 \Delta V''_\mathrm{div}(M^2).
\end{eqnarray}
Using the renormalization scheme independence of the bare parameters
we can relate the resummed and the \MSbar\ parameters at one loop
level as
\begin{eqnarray}
  \label{gapeq}
  &&\lambda -\lambda_\MSbar =  3\lambda^2 \Delta V''_\MSbar(M^2) \nn
  &&m^2 -m^2_\MSbar = \lambda \Delta V'_\MSbar(M^2) -
  \frac12\lambda^2\Phi^2 \Delta V''_\MSbar(M^2).
\end{eqnarray}
These equations determine the relation between the parameters of the
RS scheme and of the \MSbar\ scheme; but it can also be
interpreted as the formulae for how to resum diagrams in the \MSbar\ 
scheme. In fact, by the renormalization prescriptions of the
RS scheme, $\lambda$ is the complete static 4-point
function and $m^2$ is the complete mass.

The gap equations \eqref{gapeq} are derived at one loop level, and are
not necessarily useful at higher orders, that is it does not
necessarily resum any subset of diagrams.  There are arguments,
however, that help to guess the correct form. To achieve super-daisy
resummation, for example, we should use the resummed mass in the one
loop expressions: that suggest that we should always use $m^2$ in the
right hand side of \eqref{gapeq} and $\lambda$ if it is multiplied
with $\Phi^2$. Then all the possible mass resummations are done, so we
should use $\lambda_\MSbar$ for any other appearance of the coupling
constant in the expression of the mass resummation. For the explicit
coupling constants in the $\lambda$ resummation we recall that the
double scoop (two bubbles) diagram contributes $3\lambda_\MSbar^3
(\Delta V'')^2$, but if we iterate the expression for $\lambda$ as a
function of $\lambda_\MSbar$ in \eqref{gapeq}, we would get
$6\lambda_\MSbar^3 (\Delta V'')^2$. To cure this problem we should
write $\lambda\lambda_\MSbar$ instead of $\lambda^2$. So, finally, the
proposed form for the resummation reads
\begin{eqnarray}
  &&\lambda -\lambda_\MSbar =  3\lambda\lambda_\MSbar
  \Delta V''_\MSbar(M^2) \nn
  &&m^2 -m^2_\MSbar = \lambda_\MSbar \Delta V'_\MSbar(M^2) -
  \frac12\lambda\lambda_\MSbar\Phi^2 \Delta V''_\MSbar(M^2),
\end{eqnarray}
and $M^2=m^2+\frac\lambda2\Phi^2$, in agreement with 4PI suggestions
\cite{2PIformulae}.

\subsection{Computation}

From \eqref{newschemeselfen} and \eqref{eq:V''} we find $\Delta
V'(M^2)=T_B(M^2)/2$, and so we can write
\begin{eqnarray}
  && \Delta V'_\MSbar(M^2) = \frac{M^2}{32\pi^2} \ln\frac{M^2}{\mu^2} +
  \frac1{4\pi^2}\int\limits_M^\infty\! d\omega\, \sqrt{\omega^2-M^2}\,
  \;n(\omega)\nn
  && \Delta V''_\MSbar(M^2) = \frac1{32\pi^2} \ln\frac{M^2}{\mu^2} -
  \frac1{8\pi^2}\int\limits_M^\infty\!
  \frac{d\omega}{\sqrt{\omega^2-M^2}} \,n(\omega).
\end{eqnarray}
It is instructive to examine these expressions for small masses. Since
the result of the finite temperature part has $T/M$ as the only
dimensionless quantity, this is equivalent to the high temperature
approximation \cite{DolanJackiw}
\begin{equation}
  \frac1{4\pi^2}\int\limits_M^\infty\! d\omega\,\sqrt{\omega^2-M^2}\,
  \;n(\omega) = \frac{T^2}{24} -\frac{TM}{8\pi} - \frac{M^2}{32\pi^2}
  \ln\frac{M^2}{cT^2} + \c O\left(\frac{M^4}{T^2}\right),
\end{equation}
where $\ln c = 1+ 2\ln4\pi-2\gamma_E\approx4.90762$. Then, using the
notation $\mu^2= c\tilde\mu^2$, we have
\begin{eqnarray}
  && \Delta V'_\MSbar(M^2) = \frac{T^2}{24} -\frac{TM}{8\pi} +
  \frac{M^2}{32\pi^2}\ln\frac{T^2}{\tilde\mu^2} \nn 
  && \Delta V''_\MSbar(M^2) = -\frac T{16\pi M} + \frac1{32\pi^2}
  \ln\frac{T^2}{e\tilde\mu^2}.
\end{eqnarray}
Then the gap equations read
\begin{eqnarray}
  &&\frac1\lambda -\frac1{\lambda_\MSbar} =  \frac T{16\pi M} -
  \frac3{32\pi^2} \ln\frac{T^2}{e\tilde\mu^2}\\
  \label{eq:second}
  &&m^2 -m^2_\MSbar = \frac{\lambda_\MSbar T^2}{24} -
  \frac{\lambda_\MSbar T}{8\pi M}\left(M^2-\frac\lambda4\Phi^2\right)
  +\frac{\lambda_\MSbar m^2}{32\pi^2} \ln\frac{T^2}{\tilde\mu^2}
  + \frac{\lambda\lambda_\MSbar\Phi^2}{64\pi^2}.
\end{eqnarray}
The second term of the RHS of \eqref{eq:second} depends on $\Phi$ at
$\c O(\lambda^3)$ order, thus it is consistent to omit its background
dependence. The last term is the consequence of the choice of the
\MSbar\ counterterms. So we have
\begin{eqnarray}
  &&\frac1\lambda -\frac1{\lambda_\MSbar} =  \frac T{16\pi M} -
  \frac3{32\pi^2} \ln\frac{T^2}{e\tilde\mu^2}\nn
  &&m^2 -m^2_\MSbar = \frac{\lambda_\MSbar T^2}{24} -
  \frac{\lambda_\MSbar T m}{8\pi} +\frac{\lambda_\MSbar m^2}{32\pi^2}
  \ln\frac{T^2}{\tilde\mu^2} +
  \frac{\lambda\lambda_\MSbar\Phi^2}{64\pi^2}.
\end{eqnarray}
The solution therefore reads
\begin{eqnarray}
  \label{resumresult}
  && \lambda = \frac{\lambda_\MSbar}{1 - \frac{3\lambda_\MSbar}
    {32\pi^2} \ln\frac{T^2}{e\tilde\mu^2} + \frac{\lambda_\MSbar
      T}{16\pi M}} \nn 
  && m = \sqrt{ U^2 + W^2 } - W
\end{eqnarray}
where
\begin{equation}
  U^2=\frac1\kappa\left[ m_\MSbar^2 +\frac{\lambda_\MSbar T^2}{24}
    + \frac{\lambda\lambda_\MSbar\Phi^2}{64\pi^2}\right],\qquad
  W=\frac{\lambda_\MSbar T}{16\pi\kappa},\qquad \kappa =1 -
  \frac{\lambda_\MSbar}{32\pi^2}\ln\frac{T^2}{\tilde\mu^2}.
\end{equation}

The solution has some remarkable properties. First of all it is a
completely finite result, we do not need to bother with
renormalization any more. Secondly -- although coming from a high
temperature expansion -- it nicely extrapolates between the 4D and 3D
fixed points of the system. For high scales, namely,
\eqref{resumresult} can be approximated as (we set $\Phi=0$)
\begin{eqnarray}
  && \lambda = \frac{\lambda_\MSbar}{1 - \frac{3\lambda_\MSbar}
    {32\pi^2} \ln\frac{T^2}{e\tilde\mu^2}} \nn 
  && m^2 = U^2 = \frac{\displaystyle m_\MSbar^2 +\frac{\lambda_\MSbar
      T^2}{24}}{1 -
    \frac{\lambda_\MSbar}{32\pi^2}\ln\frac{T^2}{\tilde\mu^2}}.
\end{eqnarray}
The formula for $\lambda$ is the result of the non-perturbative
renormalization method of \cite{PatkosSzep}, where it was computed
by the explicit summation of bubble diagrams. 

We can realize here the running coupling and mass in the \MSbar\ 
scheme, and write
\begin{equation}
  \lambda = \lambda_\MSbar(\tilde T),\qquad m^2 = m_\MSbar^2(T)
  +\frac{\lambda_\MSbar(T) T^2}{24},
\end{equation}
where $T^2=e{\tilde T}^2$. This corresponds to a 4D behavior in the
\MSbar\ scheme.

At high temperatures the $T/m$ term takes over in the coupling
constant, and the system follows a different trajectory.  While the
mass runs in the same way as before,
\begin{equation}
  \lambda = \frac{16\pi m}{T} \approx
  16\pi\sqrt{\frac{\lambda_\MSbar(T)} {24}}.
\end{equation}

This resummation describes correctly the second order nature of the
phase transition (as opposed to the daisy resummation). Denoting
$-m_\MSbar^2 = \frac{\lambda_\MSbar}{24} T_c^2$, we find $m=0$ for
$T=T_c$. Near the critical point the solution behaves as
\begin{equation}
  m = \frac{2\pi}3(T-T_c),\qquad \lambda = \frac{16\pi m}{T_c}.
\end{equation}
The critical exponent for the mass is $\nu=1$ in this case, the fixed
point of the coupling constant is $\lambda_*=0$ (ie. it becomes
Gaussian), and the temperature dependence for the approach of this
critical point is $\lambda-\lambda_*\sim(T-T_c)^\kappa$, where
$\kappa=1$. This is exactly the behavior of the $O(N)$ model at
$N\to\infty$ \cite{Ma}.  Indeed, in this simple approximation we
resummed leading order diagrams in the large $N$ expansion, thus this
agreement is not unexpected \cite{DolanJackiw,Buchmuller}.

\section{Conclusion}
\label{sec:conclusion}

In this paper we investigated the question of to what extent the UV
and IR regularization of a system, ie. renormalization and resummation
can be reconciled. We described two methods: one is a constructive
method of how to resum higher order counterterm diagrams in order to
cure the UV divergences appearing in mass resummation in \MSbar\ 
scheme.  Similarly to the thermal mass counterterm idea we introduced
a ``compensating counterterm'' that has to be added to and subtracted
from the original Lagrangian, the two terms contributing at different
loop order. The other method is a rephrasing of this idea in the
language of the renormalization group. Here we defined a new scheme,
the resummation (RS) scheme, where the infinite part of the
counterterms cancels the UV divergences, the finite part cancels the
IR divergences of the theory at a given order. The drawback of this
scheme is that it depends explicitly on the environment, but, with
help of the renormalization group, we can project the results of the
resummation scheme to a reference scheme, eg.  \MSbar. From the point
of view of the resummation, this projection formally appears as a set
of gap equations. We demonstrated these ideas by performing coupling
constant and mass resummation in the $\Phi^4$ model.

In practice usually the problem of resummation and renormalization
comes up in two forms. The first arises when we work on a diagrammatic
basis and we compute a gap equation. At first glance the divergences
of the resummed diagrams and counterterms do not cancel each other, so
we wonder whether there is a method to consistently renormalize the
divergent gap equations. We give positive answer to this question,
proposing to use the compensating counterterm, or renormalization
group ideas described in the paper. As a rule of thumb we can use that
prescription that, for mass resummation in mass independent scheme, we
first renormalize the equation perturbatively, and we apply this
finite expression to compute the resummed parameters.

The other type of problem is when we modify our finite gap equation in
a well motivated way, hoping that it catches important higher order
effects, and we wonder whether the new form is compatible with
renormalization. The answer is again positive: any form of the gap
equation is compatible with renormalization. Technically we should
treat our gap equations as transformation equations from \MSbar\ 
scheme to a specific RS scheme. Expansion of the solution of
gap equations in terms of the coupling constants yields the finite
difference of the counterterms in the two schemes at a given order.
With the so-defined RS scheme we can do perturbation theory,
that gives the possibility to systematically improve the resummation.

In the paper we treated only resummations of the parameters of the
Lagrangian. As a future prospect we should investigate momentum
dependent resummations, too. To extend our ideas we should reinterpret
the renormalizable theories, allowing momentum dependence of the
parameters.

\section*{Acknowledgment}

The authors would like to thank Andr\'as Patk\'os and J\'anos
Polonyi for useful discussions. We acknowledge support from Hungarian
Research Fund (OTKA) under contract numbers F043465, T034980 and
T037689.

\end{document}